\documentclass[prb,twocolumn,showpacs,showkeys,amssymb]{revtex4}

\usepackage{graphicx}
\usepackage{dcolumn}
\usepackage{bm}

\begin{document}

\title{Internal transitions of negatively charged magnetoexcitons 
in quantum dots}
\author{Ricardo P{\'e}rez, Augusto Gonzalez}
\affiliation{Instituto de Cibern\'etica, Matem\'atica y F\'{\i}sica, Calle 
 E 309, Vedado, Ciudad Habana, Cuba}
\author{Jorge Mahecha}
\affiliation{Instituto de F\'{\i}sica, Universidad de Antioquia,
 AA 1226, Medell\'{\i}n, Colombia\\ and\\
 Donostia International Physics Center (DIPC), Apartado
 1072, 20080 San Sebasti{\'a}n, Spain}

\begin{abstract}
We report calculations of oscillator strengths for the far infrared 
absorption of light by the excitonic complexes $X^{n-}$ (the excess 
charge, $n$, ranging from one to four) confined in quantum dots. The 
magnetic field is varied in an interval which corresponds to ``filling 
factors'' between 2 and 3/5. Electron-hole interaction effects are seen 
in the deviations of the peak positions from the Kohn lines, and in the 
spreading of the oscillator strengths over a few final states. Transition
densities are used as an additional tool to characterize the absorption 
peaks. 
\end{abstract}


\maketitle

The lateral confinement potential of electronic quantum dots is known to 
be, to a good approximation, parabolic \cite{Hawrylak}. Thus, the far 
infrared (FIR) absorption of light by the dots exhibits a single peak, the 
Kohn mode, which position is split in a magnetic field:

\begin{equation}
\Delta E_{\pm}^{(e)}=\hbar\Omega_e\pm\frac{\hbar\omega_{ce}}{2}.
\label{eq1}
\end{equation}

\noindent
In Eq. (\ref{eq1}), $\omega_{ce}$ is the electron cyclotronic frequency, 
$\Omega_e=\sqrt{\omega_{0e}^2+\omega_{ce}^2/4}$, and $\omega_{0e}$ is the 
frequency of the lateral confinement potential. The $\pm$ signs in 
$\Delta E$ correspond, respectively, to the absorption of circularly 
polarized photons $\sigma^{\pm}$.

If a valence hole is added, but the electron-hole interaction is neglected 
for the moment, a similar mode appears for the hole:

\begin{equation}
\Delta E_{\pm}^{(h)}=\hbar\Omega_h\mp\frac{\hbar\omega_{ch}}{2}.
\label{eq2}
\end{equation}

The electron-hole interactions, however, destroys this simple picture of 
FIR absorption by spreading out the oscillator strength over a few final 
states, which positions are moved out from the Kohn lines (\ref{eq1}) and 
(\ref{eq2}). In the present paper, we study these effects by computing the 
FIR absorption of light by $X^{n-}$ complexes confined in a quantum dot. 
The excess charge, $n$, is varied from 1 to 4, that is from the so called 
trion (2 electrons, one hole) to the 5-electron, one-hole system.

Experimentally, blue shifts of the main $\sigma^+$ peaks of charged 
excitons were observed in doped quantum wells as the background electron 
density in the well increases \cite{Nickel}. The used experimental 
technique is the detection of FIR resonances by measuring changes in the 
intensity of a single photoluminescence line when a FIR radiation is 
superposed. No experimental results for quantum dots, to the best of our 
knowledge, exist, and the only theoretical computation is that of the 
trion in relatively high magnetic fields \cite{Dzyubenko}.

We employ a simple two-dimensional two-band model to describe the dot. The 
magnetic field is applied normally to the dot plane. GaAs electron- and 
heavy-hole masses, and dielectric constant are used: $m_e/m_0=0.067$, 
$m_h/m_0=0.115$, $\kappa=12.5$. The lateral confinement is 
$\hbar\omega_{0e}=1.8$ meV. It means that ``filling factors'' near one are 
reached with relatively low magnetic fields, $B=3.3$ Teslas. The latter 
was verified in a 5-electron system in which the Coulomb interactions are 
weakened by a factor 0.6 in order to account for the 
quasibidimensionality of the dot (we have in mind dots etched from a 
20-25 nm wide quantum well, for example). For holes, we choose 
$m_e\omega_{0e}=m_h\omega_{0h}$ leading to the same confinement length as 
for electrons. Calculations include 10-15 Landau levels (LLs) in the 
smallest system, the trion, and 3-4 LLs in the largest one, which 
correspond, respectively, to basis spanned by approximately 40000 and 
350000 two-dimensional Slater determinants. The estimated errors are 0.02 
meV for the excitation energies, and 0.02 for the normalized oscillator 
strengths.

\begin{figure*}[t]
\begin{center}
\includegraphics[width=.6\linewidth,angle=-90]{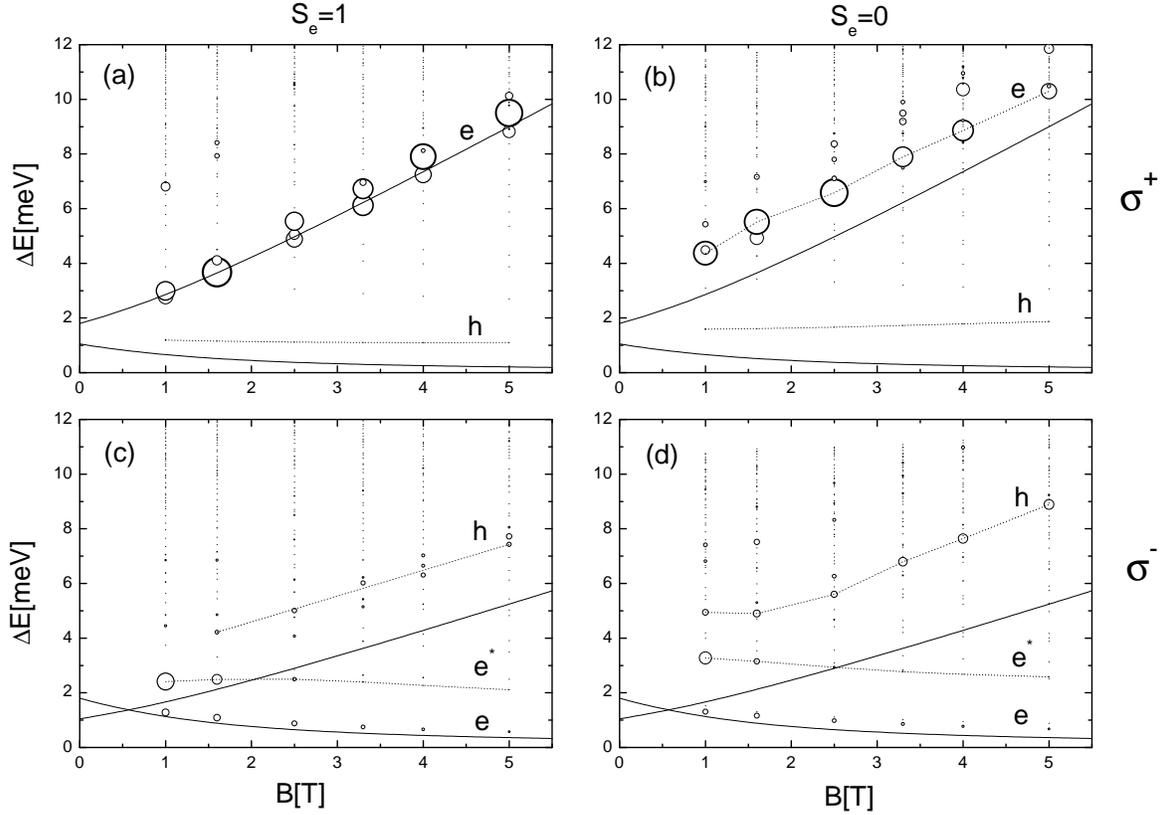}
\caption{\label{fig1} Oscillator strengths for the trion in spin triplet 
and singlet states and $\sigma^{\pm}$ polarized light.} 
\end{center} 
\end{figure*}

We summarize in Fig. \ref{fig1} the results of our calculations for the 
trion. The exact meaning of each panel in this figure is as follows. 
Let us consider, for example, the absorption of $\sigma^+$ photons by the 
trion from the lowest state with total electronic spin $S_{e}=0$ (singlet 
state), Fig. \ref{fig1} (b). According to the selection rules, 
transitions to singlet states which angular momentum projection is 
increased in one unit, $\Delta L_z=1$, take place.
The normalized oscillator strengths are represented as circles 
centered at the positions of the resonances, and with radii
proportional to the strengths. The strength normalization is based on the 
energy-weighted sum rule \cite{nuclear}:

\begin{eqnarray}
\sum_{\nu} &&\Delta 
E_{\nu}\left\{|\langle\nu|D_-|i\rangle|^2+|\langle\nu|D_+|i\rangle|^2\right\} 
\nonumber\\
&&=2 \hbar\Omega_e\left(N_e+\frac{m_e}{m_h} N_h\right),
\label{eq3}
\end{eqnarray}

\noindent
where $|i\rangle$ is the initial state, in this case the lowest singlet state, 
the $|\nu\rangle$ are all possible final states, $D_+~ (D_-)$ 
are dipole operators corresponding to the absorption of $\sigma^+~(\sigma^-)$ 
photons, $N_h=1$, and $N_e$ is the number of electrons in the dot.  
Lengths are measured in units of $l_e=\sqrt{\hbar/(m_e\Omega_e)}$, and 
energies in meV.

Qualitative features are apparent in Fig. \ref{fig1}. The most relevant ones 
are the following: (i) There are ``bands'' (i.e. series of states, 
labeled $e$ and $h$) running parallel to the Kohn lines (the latter are 
represented by solid lines). Hole bands deviate more strongly from the 
corresponding Kohn lines than electron bands. (ii) Singlet bands are 
blueshifted from triplet ($S_e=1$) bands. (iii) At low magnetic fields, the 
oscillator strengths are concentrated on $e$-bands, whereas for high $B$ 
only the $e$-bands in $\sigma^+$ polarization, and the $h$-bands in $\sigma^-$ 
polarization survive. (iv) Excited $e$-bands (labeled $e^*$) exhibiting 
high oscillator strengths and change of slopes at low magnetic fields are 
observed in $\sigma^-$ polarization.

Most of these features can be understood on simple grounds. Let us 
consider, for example, $B=0$ and let us made the rough approximation in 
which the electron-hole (e-h) interaction is $V_{e-h}=-(Ne^2/\kappa) 
/|\vec x_h-\vec X_e|$, where $\vec x_h$ and $\vec X_e$ are hole and 
electron c.m. coordinates. For simplicity, we take 
$\omega_e=\omega_h=\omega$. Then, the N-electron, one-hole hamiltonian, 
which includes interaction with the FIR radiation, takes the following form:

\begin{eqnarray}
H &=& \left\{ \frac{P^2}{2 M}+\frac{1}{2}M\omega^2 X^2 
 +(N-1)e~\vec\varepsilon (t)\cdot\vec X \right\}\nonumber\\
&+&\left\{ \frac{p^2}{2\mu}+\frac{1}{2} \mu\omega^2 x^2-
 \frac{N e^2}{\kappa x} +\frac{N(m_e+m_h)}{N m_e+m_h} 
 e~\vec\varepsilon(t)\cdot\vec x\right\}\nonumber\\
&+& H^{(e)}_{int},
\label{eq4}
\end{eqnarray}

\noindent
where we introduced $\vec x=\vec x_h-\vec X_e$, $\vec X=(m_h \vec x_h+N 
m_e \vec X_e)/M$, $M=m_h+N m_e$, $\mu=N m_e m_h/M$, $\vec\varepsilon$ is
the light polarization vector, and $H^{(e)}_{int}$ describes the internal 
motion of the electronic subsystem, which is decoupled in this 
approximation. Then, as follows from (\ref{eq4}), only the position of 
hole resonances will be affected by the e-h interactions, whereas the 
oscillator strength of the electron resonance will be stronger. That is, 
qualitatively the same as it happens in Fig. \ref{fig1} with the 
$h$-bands deviating strongly from the Kohn lines, and the $e$-bands 
concentrating the oscillator strengths at low $B$.

On the other hand, for high magnetic fields one can compute a sum rule like 
(\ref{eq3}) with the inclusion of only single-particle states in the first 
LL. The r.h.s. of this sum rule is given by $2 \hbar (\Omega_e-
\hbar\omega_{ce}/2)(N_e+N_h m_e/m_h)$. It means that transitions within 
the first LL are suppressed as $B$ increases. Only $e$-bands in $\sigma^+$ 
polarization and $h$-bands in $\sigma^-$ polarization have nonvanishing 
strengths in this limit.

\begin{figure*}[ht]
\begin{center}
\includegraphics[width=.6\linewidth,angle=-90]{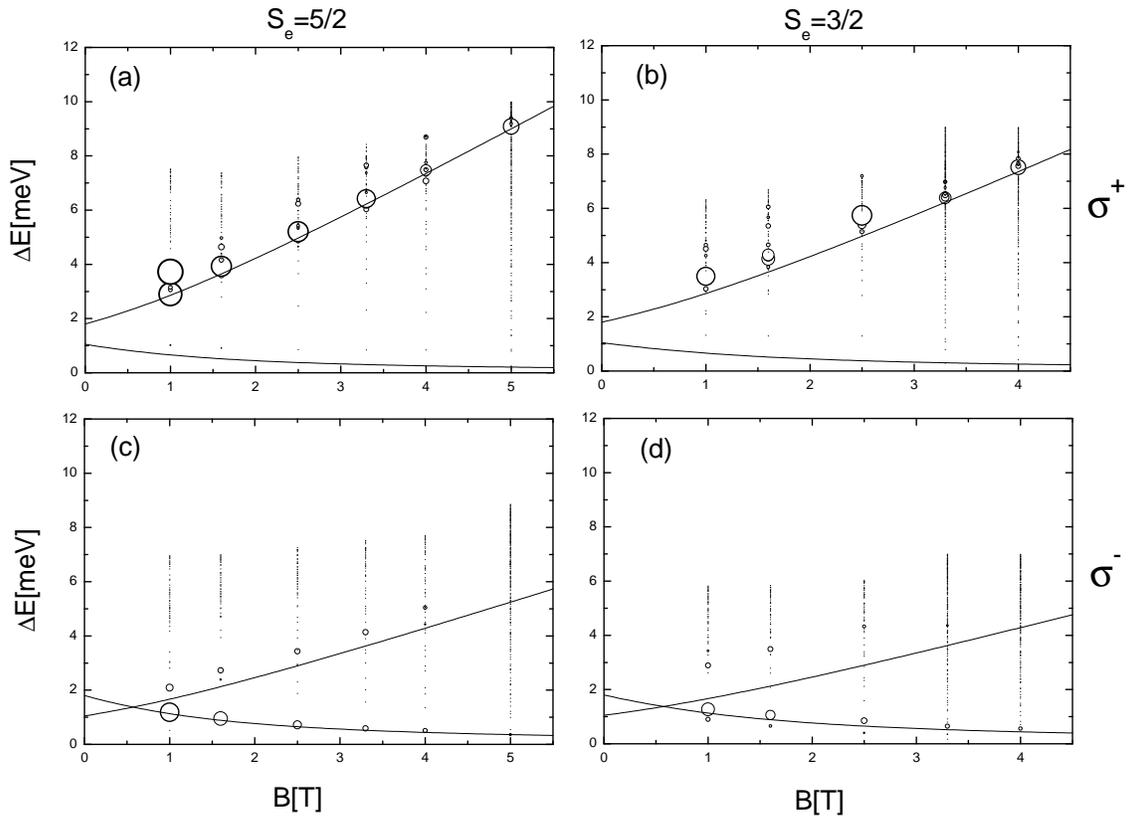}
\caption{\label{fig2} Oscillator strengths for the $X^{4-}$ complex.
See explanation in the main text.}
\end{center}
\end{figure*}

The fact about the blueshift of singlet bands has also a simple 
interpretation. Indeed, in singlet states, the electronic cloud is more 
compactly localized around the hole than in triplet states. Consequently, 
e-h interactions are stronger and collective motion of electrons, for 
example, will have higher excitation energies. With respect to the 
excited, $e^*$-bands, and the change of slopes, we ascribe these features 
to mode mixing, i.e. to $e$- and $h$-band admixtures. Oscillations of the 
electron cloud in counterphase with the hole cloud are the dominant 
resonances in the FIR absorption of neutral systems \cite{GDR,biexciton}.
In charged excitons, the Kohn modes dominate, but there are states 
resembling these counterphase oscillations, which can be viewed as mode 
admixtures. 

To conclude the analysis of the trion, we stress that for bands which 
energy increases with $B$, the number of levels below the band also 
increases. It means that the ``transfer'' of oscillator strength between 
resonances in a band goes through ``level collisions'' as the magnetic 
field varies. A similar situation was reported for the biexciton 
\cite{biexciton}. 

Excitonic complexes with excess charge $n>1$ exhibit the same qualitative 
features (i--iv) mentioned above for the trion. There are, in addition, new 
aspects which can be summarized as follows: (I) These complexes undergo 
ground-state (gs) rearrangements as $B$ is varied between 1 and 5 Teslas. 
Thus, resonances in a band may correspond to transitions which start at 
different initial states. On the other hand, as a function of $n$: (II) The 
main $e$-band in $\sigma^+$ polarization first experiences a redshift (for 
$n=3$ and spin-polarized states it is already below the Kohn line), and 
then for $n=4$ (and presumably for larger values also) is blueshifted. 
(III) The main $h$-band in $\sigma^-$ polarization moves down in energy from 
$n=1$ up to $n=4$ staying, however, above the Kohn line. (IV) The $e$- or 
$e^*$-bands in $\sigma^-$ polarization become redshifted up to $n=4$.

We show in Fig. \ref{fig2} the results for the FIR absorption by the 
$X^{4-}$ complex in $S_e=5/2$ (spin-polarized) and $S_e=3/2$ states, and 
$\sigma^{\pm}$ light polarizations. The main $e$- and 
$h$-bands, and even the $e^*$-bands, the blueshift of $S_e=3/2$ bands with 
respect to $S_e=5/2$ bands, etc. are observed.

\begin{figure*}[ht]
\begin{center}
\includegraphics[width=.6\linewidth,angle=-90]{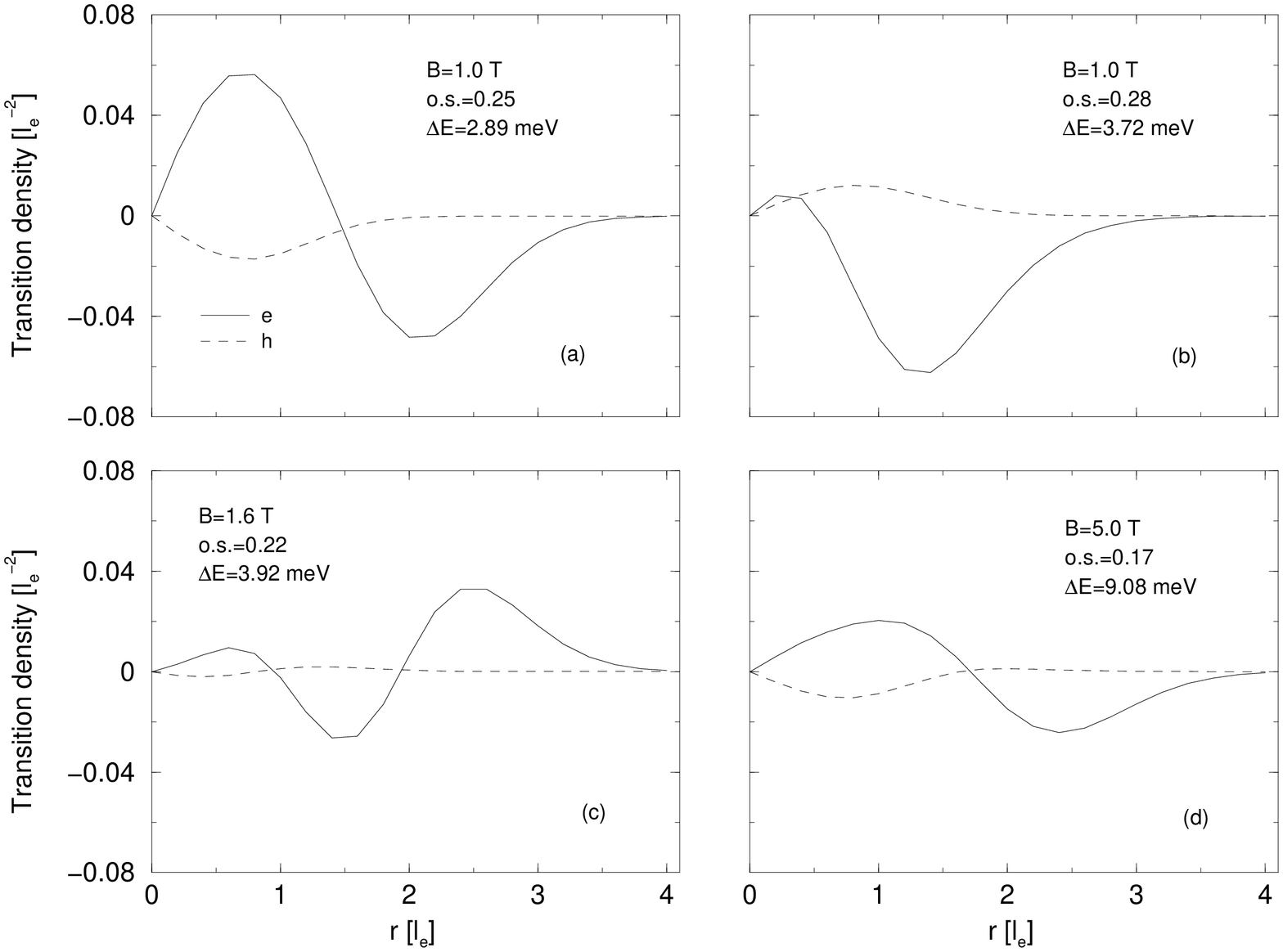}
\caption{\label{fig3} Transition densities for the $X^{4-}$ complex in 
spin-polarized states and $\sigma^+$ light polarization. The excitation 
energies and oscillator strengths of the resonances are indicated.}
\end{center}
\end{figure*}

Let us consider the question about gs rearrangements and their 
consequences. In the trion, no such phenomenon exists. The lowest triplet 
state is always a state with $L_z=-1$, and the lowest singlet state is a 
state with $L_z=0$. Mode mixing proceeds continuously, leading to change 
of band slopes at low magnetic fields, etc. In the larger complexes, gs 
rearrangements may lead to discontinous variations of the properties of a 
resonance along a band, because a given oscillation mode may not be 
supported by the new ground state. As a consequence, the oscillator 
strength is transferred to a state with different properties. 

This effect is seen in the transition densities \cite{nuclear}: 
$\langle\nu|\hat\rho(\vec r)| i\rangle$, where $\hat\rho(\vec r)$ is the 
one-particle density operator. Transition densities enter the expression 
for the time-dependent density, which is the result of a perturbation (the 
external FIR radiation field) around the initial state:

\begin{eqnarray}
\rho(\vec r,t)&=&\rho_i(\vec r)+\sum_{\nu}\left\{\langle 
 i|\hat\rho(\vec r)|\nu\rangle ~C_{\nu} 
 e^{-iE_{\nu}t/\hbar}\right.\nonumber\\
&+& \left. \langle \nu|\hat\rho(\vec r)|i\rangle ~C_{\nu}^* 
 e^{iE_{\nu}t/\hbar}\right\}+\dots
\end{eqnarray}

\noindent
Where $C_{\nu}$ are the coefficients of the spectral decomposition of
the perturbed state.
In Fig. \ref{fig3}a and b, we present the two leading $e$-resonances at 
$B=1$ Tesla in the spin-polarized $S_e=5/2$ system, and $\sigma^+$ 
polarization. Both resonances are characterized by dominant electronic 
oscillations and counterphase oscillations of electrons and holes. At 
$B=1.6$ Teslas we observe the first gs rearrangement. The gs angular 
momentum jumps from $L_z=-5$ to $L_z=-10$. The transition density of the 
leading $e$-resonance is given in Fig. \ref{fig3}c, showing an additional 
nodal line. The qualitative form of the transition density for the leading 
$e$-resonance persists up to $B=4$ Teslas. At $B=5$ Teslas a new gs 
rearrangement takes place, in which the momentum jumps to $L_z=-14$. The 
leading $e$-resonance is shown in Fig. \ref{fig3}d. The qualitative 
behavior changed again. The conclusion is that, in the larger complexes, 
resonances along a band may correspond to transitions from different 
initial states, and represent qualitatively different oscillation modes 
against these ground states.

Finally, let us stress that the shift of resonance bands as a function of 
the excess charge $n$ is a result of Coulomb interactions in these 
few-body complexes. Unlike the quantum-well case \cite{Nickel}, we obtain 
a redshift of the main $e$-band in $\sigma^+$ polarization for $n=3$. 
For $n=4$, the band moves again to higher excitation energies. A 
qualitative understanding of the behavior of these bands is still lacking.

\begin{acknowledgments}
The authors acknowledge support from the Commitee for Research 
of the Universidad de Antioquia (CODI) and the Colombian Institute for 
Science and Technology (COLCIENCIAS). JM gratefully acknowledges help and 
support by the Donostia International Physics Center (DIPC).
\end{acknowledgments}

\end{document}